\documentclass[a4paper,superscriptaddress,citeautoscript,twocolumn]{revtex4-2}

\usepackage[T1]{fontenc}
\usepackage{times}
\usepackage{amsmath,amssymb}
\usepackage[pdftex]{graphicx,color}
\usepackage{epstopdf}
\usepackage[a4paper,body={18cm,25.5cm},top=2cm]{geometry}
\usepackage[english]{babel}

\begin{document}

\title{Parametrically-driven temporal cavity solitons in a bichromatically-driven pure Kerr resonator}

\author{Miriam Leonhardt}
\thanks{These authors contributed equally to this work.}
\affiliation{Department of Physics, University of Auckland, Auckland 1010, New Zealand}
\author{David Paligora}
\thanks{These authors contributed equally to this work.}
\affiliation{Department of Physics, University of Auckland, Auckland 1010, New Zealand}
\author{Nicolas Englebert}
\affiliation{Service OPERA-Photonique, Universit\'e libre de Bruxelles (U.L.B.), 50 Avenue F. D. Roosevelt, CP 194/5, B-1050 Brussels, Belgium}
\author{Fran\c{c}ois Leo}
\affiliation{Service OPERA-Photonique, Universit\'e libre de Bruxelles (U.L.B.), 50 Avenue F. D. Roosevelt, CP 194/5, B-1050 Brussels, Belgium}
\author{Julien Fatome}
\affiliation{Department of Physics, University of Auckland, Auckland 1010, New Zealand}
\affiliation{Laboratoire Interdisciplinaire Carnot de Bourgogne, UMR 6303 CNRS Universit\'e Bourgogne-Franche-Comt\'e, Dijon, France}
\author{Miro Erkintalo}
\email{m.erkintalo@auckland.ac.nz}
\affiliation{Department of Physics, University of Auckland, Auckland 1010, New Zealand}
\affiliation{The Dodd-Walls Centre for Photonic and Quantum Technologies, New Zealand}

\begin{abstract}
  \noindent Temporal cavity solitons (CSs) are pulses of light that can persist endlessly in dispersive, nonlinear optical resonators. They have been extensively studied in the context of resonators with purely cubic (Kerr-type) nonlinearity that are externally-driven with a monochromatic continuous wave laser -- in such systems, the solitons manifest themselves as unique attractors whose carrier frequency coincides with that of the external driving field. Recent experiments have, however, shown that a qualitatively different type of temporal CS can arise via parametric down-conversion in resonators with simultaneous quadratic and cubic nonlinearity. In contrast to conventional CSs in pure-Kerr resonators, these \emph{parametrically-driven cavity solitons} come in two different flavours with opposite phases, and they are spectrally centred at half of the frequency of the driving field. Here, we theoretically and numerically show that, under conditions of bichromatic driving, such parametrically-driven CSs can also arise in dispersive resonators with pure Kerr nonlinearity. In this case, the solitons arise through parametric four-wave mixing, come with two distinct phases, and have a carrier frequency in between the two external driving fields. We show that, when all waves are resonant, the solitons can experience long-range interactions due to their back-action on the intracavity fields at the pump frequencies, and we discuss the parameter requirements for the solitons' existence. Besides underlining the possibility of exciting a new type of cavity soliton in dispersive Kerr cavities, our work advances the theoretical modeling of resonators that are coherently-driven with polychromatic fields.

\end{abstract}

\maketitle
\section{Introduction}

The injection of monochromatic continuous wave (CW) laser light into dispersive optical resonators with purely Kerr-type nonlinearity can lead to the generation of localized dissipative structures know as temporal Kerr cavity solitons (CSs)~\cite{leo_temporal_2010, jang_ultraweak_2013, kippenberg_dissipative_2018}. These CSs correspond to ultrashort pulses of light that can persist within the resonator, maintaining constant shape and energy through a double-balance between group-velocity dispersion and Kerr nonlinearity on the one hand, and dissipation and external driving on the other hand~\cite{wabnitz_suppression_1993}. While first observed and studied in macroscopic optical fiber ring resonators~\cite{leo_temporal_2010}, CSs have attracted particular attention in the context of monolithic Kerr microresonators~\cite{kippenberg_dissipative_2018}, where they underpin the generation of coherent and broadband optical frequency combs~\cite{herr_temporal_2014,brasch_photonic_2016,pasquazi_micro-combs_2018}.

The conventional CSs that manifest themselves in resonators with pure Kerr nonlinearity sit atop a CW background, and they gain their energy through four-wave interactions with that background~\cite{leo_temporal_2010}. In the frequency domain,  the solitons are to first order centred around the frequency of the external CW laser that drives the resonator~\cite{herr_temporal_2014} . They are (barring some special exceptions~\cite{nielsen_coexistence_2019,anderson_coexistence_2017,hansson_frequency_2015,xu_spontaneous_2021,lucas_spatial_2018}) unique attracting states: except for trivial time translations, all the CSs that exist for given system parameters are identical. Interestingly, recent experiments have revealed that qualitatively different types of temporal CSs -- that do not share the aforementioned characteristics -- can exist in resonators that display a quadratic nonlinearity in addition to a Kerr-type nonlinearity. In particular, Englebert et al. have experimentally demonstrated that an all-fibre optical parametric oscillator driven at  $2\omega_0$ can support CSs at $\omega_0$~\cite{englebert_parametrically_2021}; in this configuration, the solitons are \emph{parametrically-driven} through the quadratic down-conversion of the externally-injected field. The CSs that result from such parametric driving display a hyperbolic secant amplitude profile akin to those manifesting themselves in monochromatically-driven Kerr-only resonators, but the two types of solitons nonetheless exhibit significant differences. Importantly, parametrically-driven CSs are spectrally separated from the driving frequency ($\omega_0$ versus $2\omega_0$ in the configuration used in~\cite{englebert_parametrically_2021}), and they come in two distinct states with opposite phases. These traits render parametrically-driven CSs of interest for an altogether new range of applications.

In theory, parametrically-driven CSs can be supported by any nonlinearity that provides phase-sensitive amplification~\cite{mecozzi_long-term_1994}. While ref.~\cite{englebert_parametrically_2021} realised suitable conditions using quadratic down-conversion, it is well-known that analogous phase-sensitive amplification can also be realised in pure Kerr resonators using appropriate driving configurations~\cite{mecozzi_long-term_1994,agrawal_nonlinear_nodate,radic_two-pump_2003, okawachi_dual-pumped_2015, Andrekson_fiber-based_2020}. In fact, there has been considerable recent interest in leveraging phase-sensitive four-wave mixing to realise bi-phase (non-solitonic) states in pure Kerr resonators driven by two lasers with different carrier frequencies; such systems have allowed for the realisation of novel random number generators~\cite{takesue_10_2016,okawachi_quantum_2016,okawachi_dynamic_2021} as well as coherent optical Ising machines~\cite{inagaki_Large-scale_2016, mohseni_ising_2022}. A natural question that arises is: is it possible to generate parametrically-driven CSs in Kerr-only resonators with bichromatic driving? Whilst this question has been previously explored in the context of \emph{diffractive} Kerr-only resonators~\cite{de_valcarcel_phase-bistable_2013}, the presence of \emph{dispersion} substantially changes the physics of the problem. The impact of bichromatic driving in the dynamics of conventional Kerr CSs has also been considered~\cite{hansson_bichromatically_2014, ceoldo_multiple_2016, zhang_spectral_2020,moille_ultra-broadband_2021, qureshi_soliton_2021, taheri_all-optical_2022}, but to our knowledge, no work has yet explored the possibility of using the scheme to generate parametrically-driven CSs in pure Kerr resonators.

\begin{figure}[!t]
 \centering
  \includegraphics[width = \columnwidth, clip=true]{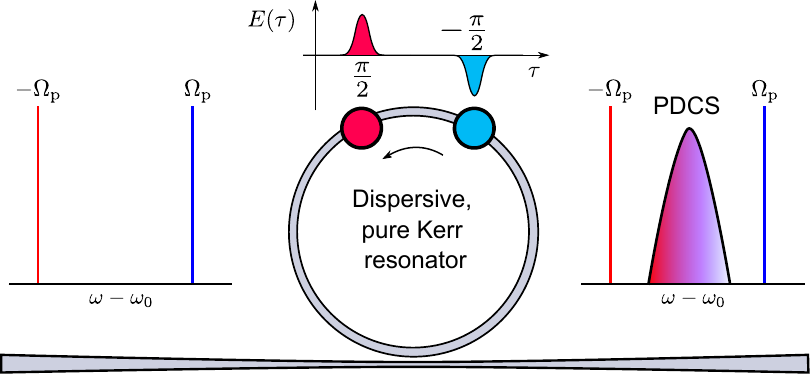}
 \caption{Schematic illustration of parametrically-driven CS (PDCS) generation in a dispersive, pure Kerr resonator driven with two monochromatic fields detuned by $\pm\Omega_\mathrm{p}$ from the signal frequency $\omega_0$. The PDCSs come in two opposite phases and are spectrally centred in between the injected driving fields at $\omega_0$.}
 \label{fig1}
\end{figure}

In this article, we theoretically and numerically show that a dispersive resonator with pure Kerr nonlinearity can support parametrically-driven CSs in the presence of bichromatic driving. Specifically, we show that a signal field with carrier frequency in between two spectrally-separated driving fields obeys the damped, parametrically-driven nonlinear Schr\"odinger equation that admits soliton solutions, and we numerically confirm that such solitons can be realised under appropriate conditions. We show that the parametrically-driven cavity solitons (PDCSs) come in two different phases, and that coexisting solitons can experience long-range interactions under conditions where all interacting waves are resonant. We also discuss the parameter requirements that underpin the solitons' existence, as well as pathways to experimental implementations. Besides unveiling a new class of temporal CSs in coherently-driven Kerr resonators, our analyses advance the modelling of resonators subject to polychromatic driving.

\section{Models}

We begin by discussing the theoretical modelling of bichromatically-driven Kerr resonators. Our starting point is a polychromatic Ikeda-like map, which we will use to derive an extended mean-field Lugiato-Lefever equation that has been used in previous studies~\cite{hansson_bichromatically_2014, taheri_optical_2017, zhang_spectral_2020,moille_ultra-broadband_2021, qureshi_soliton_2021, taheri_all-optical_2022}.  To this end, we consider a Kerr resonator made out of a dispersive waveguide [with length $L$ and propagation constant $\beta(\omega)$] that is driven with two coherent fields with angular frequencies $\omega_\pm = \omega_0 \pm \Omega_\mathrm{p}$ [see Fig.~\ref{fig1}]. The evolution of the electric field envelope (referenced against the carrier frequency $\omega_0$ of the parametric signal) during the $m$th transit around the resonator is governed by the generalized nonlinear Schr\"odinger equation:
\begin{equation}
\frac{\partial E^{(m)}(z,\tau)}{\partial z} = i\hat{D}_\mathrm{S}\left(i\frac{\partial}{\partial\tau}\right)E^{(m)} + i\gamma |E^{(m)}|E^{(m)}. \label{Seq}
\end{equation}
Here $z$ is a coordinate along the waveguide that forms the resonator, $\tau$ is time in a reference frame that moves at the group-velocity of light at $\omega_0$, $\gamma$ is the Kerr nonlinearity coefficient and the dispersion operator
\begin{equation}
\hat{D}_\mathrm{S}\left(i\frac{\partial}{\partial\tau}\right) = \sum_{k\geq 2} \frac{\beta_k}{k!}\left(i\frac{\partial}{\partial\tau}\right)^k,
\label{dispop}
\end{equation}
with $\beta_k = d^k\beta/d\omega^k|_{\omega_0}$ the Taylor series expansion coefficients of $\beta(\omega)$ around $\omega_0$. Note that the single electric field envelope $E^{(m)}(z,\tau)$ contains all the frequency components pertinent to the nonlinear interactions, including the fields at the pump frequencies $\omega_0\pm\Omega_\mathrm{p}$ and the signal frequency at $\omega_0$~\cite{hansson_single_2016}.

The Ikeda map consists of Eq.~\eqref{Seq} together with a boundary equation that describes the coupling of light into the resonator. Considering bichromatic driving with frequencies $\omega_0 \pm \Omega_\mathrm{p}$, the boundary equation reads [see Appendix~\ref{apA}]:
\begin{align}
E^{(m+1)}(0,\tau) &= \sqrt{1-2\alpha} E^{(m)}(L,\tau)e^{-i\delta_0} \nonumber \\
 &+ \sqrt{\theta_+}E_\mathrm{in,+} e^{-i\Omega_\mathrm{p}\tau + imb_+} \nonumber \\
 &+ \sqrt{\theta_-}E_\mathrm{in,-} e^{i\Omega_\mathrm{p}\tau + imb_-}. \label{Boundary}
\end{align}
Here $\alpha$ is half of the fraction of power dissipated by the intracavity field over one round trip, $\delta_0 = 2\pi k - \beta(\omega_0) L$ is the linear phase detuning of the reference frequency $\omega_0$ from the closest cavity resonance (with order $k$), and $E_\mathrm{in,\pm}$ are the complex amplitudes of the driving fields at $\omega_\pm = \omega_0 \pm \Omega_\mathrm{p}$, respectively, with $\theta_\pm$ the corresponding power transmission coefficients that describe the coupling of the driving fields into the resonator. The coefficients $b_\pm$ allow us to introduce the phase detunings $\delta_\pm$ that describe the detunings of the pump frequencies from the cavity resonances closest to them:
\begin{equation}
b_\pm = \delta_\pm -\delta_0 +\hat{D}_\mathrm{S}(\pm\Omega_\mathrm{p})L. \label{bs}
\end{equation}
Before proceeding, we note that, in our specific configuration, only two out of the three detuning terms introduced above ($\delta_0$ and $\delta_\pm$) are independent. This is because the central signal frequency is completely determined by the pump frequencies viz. $\omega_0 = (\omega_+ + \omega_-)/2$; therefore, the signal detuning $\delta_0$ can be written in terms of the pump detunings $\delta_\pm$ as [see Appendix~\ref{apA2}]:
\begin{equation}
\delta_0 = \frac{\delta_+ + \delta_- + L[\hat{D}_\mathrm{S}(\Omega_\mathrm{p})+\hat{D}_\mathrm{S}(-\Omega_\mathrm{p})]}{2}. \label{d0}
\end{equation}
Substituting this expression for $\delta_0$ into Eq.~\eqref{bs} yields $b_\pm = \pm b$, where
\begin{equation}
b = \frac{\delta_+ - \delta_- + L[\hat{D}_\mathrm{S}(\Omega_\mathrm{p})-\hat{D}_\mathrm{S}(-\Omega_\mathrm{p})]}{2}. \label{bcoef}
\end{equation}

Under the assumption that the intracavity envelope $E^{(m)}(z,\tau)$ evolves slowly over a single round trip (i.e., the cavity has a high finesse, and the linear and nonlinear phase shifts are all small), the Ikeda-like map described above can be averaged into the generalized Lugiato-Lefever mean-field equation similar to the one used, e.g., in refs.~\cite{zhang_spectral_2020,moille_ultra-broadband_2021, qureshi_soliton_2021, taheri_all-optical_2022}. We write the equation in normalized form as [see Appendix~\ref{apB}]:
\begin{align}
\frac{\partial E(t,\tau)}{\partial t} &= \left[-1 + i(|E|^2 -\Delta_0)+ i\hat{D}\left(i\frac{\partial}{\partial\tau}\right)\right]E \label{LLN} \\
 &+ S_+ e^{-i\Omega_\mathrm{p}\tau + iat} \nonumber + S_- e^{i\Omega_\mathrm{p}\tau - iat}.  \nonumber
\end{align}
Here $t$ is a slow time variable that describes the evolution of the intracavity field over consecutive round trips (and is thus directly related to the index $m$ of the Ikeda-like map), \mbox{$S_\pm = E_\mathrm{in,\pm} \sqrt{\gamma L \theta_\pm/\alpha^3}$} are the normalized strengths of the driving fields, $\Delta_0 = \delta_0/\alpha$ is the normalized detuning of the signal field, and the normalized dispersion operator $\hat{D}$ is defined as Eq.~\eqref{dispop} but with normalized Taylor series coefficients $\beta_k\rightarrow d_k = [2\alpha/(|\beta_2|L)]^{k/2}\beta_kL/\alpha$. Finally, the coefficient
\begin{equation}
a = \frac{b}{\alpha} = \frac{\Delta_+ - \Delta_- + [\hat{D}(\Omega_\mathrm{p})-\hat{D}(-\Omega_\mathrm{p})]}{2}, \label{acoef}
\end{equation}
where $\Delta_\pm = \delta_\pm/\alpha$ are the normalized detunings of the external driving fields. To avoid notational clutter, we use the symbol $\Omega_\mathrm{p}$ to represent pump frequency shifts both in our dimensional and normalized equations.

It is worth noting that, in the special case where the driving fields have identical amplitudes ($S_+ = S_- = S_0/2$), the external driving terms in Eq.~\eqref{LLN} can be combined to yield
\begin{equation}
S(t,\tau) = S_0\cos(\Omega_\mathrm{p}\tau - at). \label{cosdrive}
\end{equation}
This expression reveals that the coefficient $a$ represents a desynchronization between the envelope formed from the superposition of the driving fields and an integer fraction of the cavity round trip time [see ref.~\cite{englebert_bloch_2021} and Appendix~\ref{apA2}]. Interestingly, however, this desynchronization plays no role at all in the PDCS dynamics. We also remark that, while previous studies have shown that sinusoidally-driven dispersive Kerr resonators can admit localized soliton states~\cite{hansson_bichromatically_2014, ceoldo_multiple_2016}, the focus has been on comparatively small modulation frequencies. In that case, the solitons correspond to conventional Kerr CSs that are temporally confined to within a single period of the driving modulation. In stark contrast, here we find that PDCSs arise in the qualitatively different limit of large pump frequency spacings, with their duration being significantly larger than a single period of the modulation defined by the pump fields. We emphasize also that the special case $S_+ = S_- = S_0/2$ is \emph{not} a pre-requisite for PDCS existence, but is described above simply to highlight the physical interpretation of the coefficient $a$.

\section{PDCS theory}

Most of the numerical simulation results presented below have been obtained using the full Ikeda map described by Eqs.~\eqref{Seq} and~\eqref{Boundary}, but are presented in normalized units for the sake of generality (the Ikeda simulations consider a resonator finesse of $\mathcal{F}=100$ to define the variable $\alpha = \pi/\mathcal{F}$). While the mean-field Eq.~\eqref{LLN} yields almost identical results (as will be illustrated), the Ikeda map is ultimately more general and faster to integrate numerically when the desynchronization coefficient $a\neq 0$ (such that Eq.~\eqref{LLN} becomes non-autonomous). However, the mean-field Eq.~\eqref{LLN} allows us to more transparently lay the theoretical foundations of the concept. To this end, we make the assumption that the intracavity fields $E_\pm$ at the pump frequencies are homogeneous and stationary. (Note: this assumption will not be used in any of our simulations.) We then substitute the ansatz
\begin{align}
E(t,\tau) &= E_0(t,\tau) \label{anz}\\
&+ E_+e^{-i\Omega_\mathrm{p}\tau + ia t} + E_-e^{i\Omega_\mathrm{p}\tau - ia t}  \nonumber
\end{align}
into Eq.~\eqref{LLN}, and  assume that the (soliton) spectrum around the signal frequency (the Fourier transform of $E_0(t,\tau)$) does not exhibit significant overlap with the pump frequencies. This allows to separate terms that oscillate with different frequencies, yielding the following equation for the signal field:
\begin{align}
\frac{\partial E_0(t,\tau)}{\partial t} &=\left[-1 + i(|E_0|^2 -\Delta_\mathrm{eff}) + i\hat{D}\left(i\frac{\partial}{\partial\tau}\right) \right]E_0 \nonumber \\
&+ 2iE_+E_- E_0^\ast, \label{PDNLSE}
\end{align}
where the effective detuning $\Delta_\mathrm{eff} = \Delta_0 - 2(Y_+ + Y_-)$ with $Y_\pm=|E_\pm|^2$ includes both linear and nonlinear (cross-phase modulation) phase shifts. Equation~\eqref{PDNLSE} has the precise form of the parametrically-driven nonlinear Schr\"odinger equation~\cite{bondila_topography_1995} with effective detuning $\Delta_\mathrm{eff}$ and parametric driving coefficient $\mu = 2iE_+E_-$. (Note that the desynchronization coefficient $a$ cancels in the product $E_+E_-$ which is why that term does not affect the PDCS dynamics.) Accordingly, assuming that the resonator group-velocity dispersion is anomalous at the signal frequency ($\beta_2 < 0$), the equation admits exact (parametrically-driven) soliton solutions of the form~\cite{englebert_parametrically_2021}:
\begin{equation}
E_0(\tau) = \sqrt{2}\beta \text{sech}(\beta\tau)e^{i(\phi + \theta)}, \label{PDsol}
\end{equation}
where $\cos(2\phi) = 1/|\mu|$, $\beta = \sqrt{\Delta_\mathrm{eff}+|\mu|\sin(2\phi)}$, and \mbox{$\theta = \text{arg}[iE_+E_-]$}.

The soliton solution given by Eq.~\eqref{PDsol} is strictly valid only under the assumption that the intracavity fields $E_\pm$ at the pump frequencies are homogeneous. Although this assumption does not hold in general, judicious choice of the pump parameters and resonator dispersion can provide conditions under which the assumption holds to a sufficient degree. First, if either of the pump frequencies $\pm\Omega_\mathrm{p}$ experiences anomalous group-velocity dispersion ($\hat{D}_2(\pm \Omega_\mathrm{p})<0$ where $\hat{D}_2=d^2\hat{D}/d\Omega^2$), then the corresponding intracavity intensity $Y_\pm$ must be below the threshold of modulation instability that would otherwise break the homogeneity of the field~\cite{coen_universal_2013}. Second, the temporal walk-off between the signal field $E_0$ and the fields at the pump frequencies ($\hat{D}_1(\pm\Omega_\mathrm{p})$ where $\hat{D}_1=d\hat{D}/d\Omega$) must be sufficiently large so as to mitigate pump depletion in the vicinity of the soliton that would otherwise break the homogeneity of the fields $E_\pm$. In effect, this latter requirement implies that the soliton has negligible effect on the intracavity fields at the pump frequencies.

An additional requirement on the system parameters is imposed by the fact that dispersion links the linear detunings experienced by all of the waves. Recalling Eq.~\eqref{d0}, the effective detuning of the signal field can be written as
\begin{equation}
\Delta_\mathrm{eff} = \frac{\Delta_+ + \Delta_- + \hat{D}(\Omega_\mathrm{p}) + \hat{D}(-\Omega_\mathrm{p})}{2} - 2\left(Y_+ + Y_-\right). \label{PM}
\end{equation}
The value of $\Delta_\mathrm{eff}$ must naturally be such that soliton existence is possible, which for given dispersion conditions significantly restricts the applicable pump frequencies $\Omega_\mathrm{p}$ to a narrow range. Considering typical parameters, $\Delta_\mathrm{eff}$ and $|\mu| = 2\sqrt{Y_+Y_-}$ are of the order of unity for solitons to exist~\cite{englebert_parametrically_2021}, while the detunings $\Delta_\pm$ can be assumed small to ensure that sufficient intracavity powers $Y_\pm$ can be attained without excessive driving powers $X_\pm = |S_\pm|^2$. This implies, then, that the pump frequency shift must satisfy $[\hat{D}(\Omega_\mathrm{p}) + \hat{D}(-\Omega_\mathrm{p})]/2 = \hat{D}_e(\Omega_\mathrm{p})\approx 0$, where $\hat{D}_e(\Omega_\mathrm{p})$ only contains the even-order dispersion terms. It is worth noting that this condition [$\hat{D}_e(\Omega_\mathrm{p})\approx 0$] is congruent with the linear phase-matching condition for degenerate four-wave mixing in Kerr resonators~\cite{sayson_widely_2017, sayson_octave-spanning_2019}, which is unsurprising given that the parametric interaction driving the signal field $E_0(t,\tau)$ is precisely that process.

All of the conditions above can be met when the resonator has non-negligible fourth-order dispersion $d_4$~\cite{sayson_widely_2017, sayson_octave-spanning_2019}. As an example, for $\hat{D}(\Omega) = d_2\Omega^2/2 + d_4\Omega^4/24$, the condition $\hat{D}_e(\Omega_\mathrm{p}) \approx 0$ yields a single non-trivial pump frequency shift $\Omega_\mathrm{p} = \sqrt{12|d_2|/d_4}$; this frequency shift is associated with normal group-velocity dispersion at both pump frequencies (which prohibits modulation instabilities) as well as non-zero temporal walk-off with $\hat{D}_1(\Omega_\pm)\rightarrow \pm \infty$ as $d_4\rightarrow 0$. We emphasize that resonators with suitable dispersion already exist, as demonstrated by observations of large-frequency-shift parametric oscillation phase-matched via fourth-order dispersion~\cite{sayson_widely_2017, sayson_octave-spanning_2019, fujii_third-harmonic_2017, lu_efficient_2019, lu_milliwatt-threshold_2019, li_experimental_2020}.

When the conditions described above are met, portions of the intracavity fields $E_\pm$ at the pump frequencies can be approximated to be homogeneous. Because the action of the soliton on the intracavity fields at the pump frequencies can be assumed negligible (thanks to dispersive walk-off), in (quasi) steady-state those fields will have complex amplitudes $E_\pm \approx S_\pm/[1-i(Y_\pm + 2Y_\mp - \Delta_\pm)]$, with the intensities $Y_\pm = |E_\pm|^2$ satisfying the following coupled algebraic equations (see Appendix~\ref{apC} and refs.~\cite{haelterman_polarization_1994, del_bino_symmetry_2017, garbin_asymmetric_2020}):
\begin{equation}
X_\pm = \left[1 + (Y_\pm + 2Y_\mp - \Delta_\pm)^2 \right]Y_\pm. \label{Xs}
\end{equation}
By appropriately choosing the driving parameters $\Delta_\pm$ and $X_\pm$, it is possible to achieve intracavity powers $Y_\pm$ that provide sufficiently large parametric driving strength $|\mu| = 2\sqrt{Y_+Y_-}$ to permit parametrically-driven solitons. Here it is worth noting that, while symmetric states satisfying $Y_+ = Y_-$ can always be found for symmetric parameters ($X_+ = X_-$ and $\Delta_+ = \Delta_-$), the coupled Eqs.~\eqref{Xs} are known to display spontaneous symmetry breaking~\cite{haelterman_polarization_1994, del_bino_symmetry_2017, garbin_asymmetric_2020}, which can make the symmetric states unstable. Once the detunings $\Delta_\pm$ and intracavity intensities $Y_\pm$ are known, one can solve for the pump frequency shift $\Omega_\mathrm{p}$ from Eq.~\eqref{PM}.

\section{Simulations}

\begin{figure}[!t]
 \centering
  \includegraphics[width = \columnwidth, clip=true]{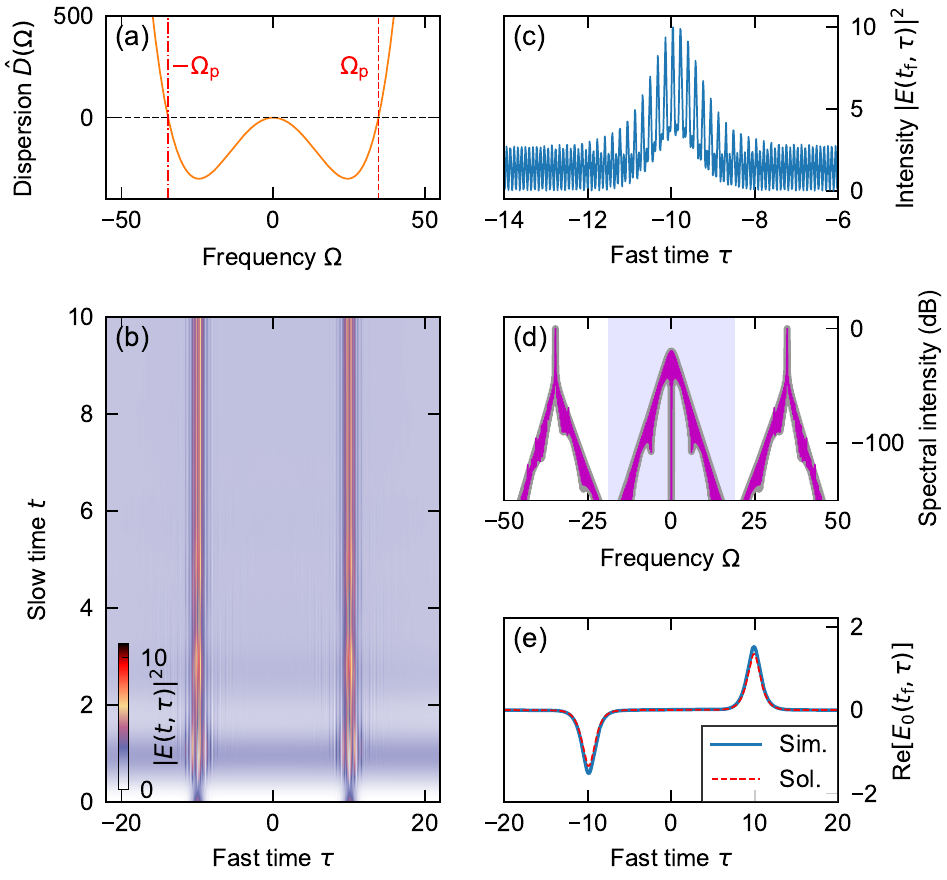}
 \caption{Simulation results illustrating PDCS generation in a pure Kerr resonator. (a) Dispersion $\hat{D}(\Omega) = d_2\Omega^2/2 + d_4\Omega^4/24$ with $d_2 = -2$ and $d_4 = 0.02$ used in the simulations. The dash-dotted vertical lines indicate the pump frequency shifts with $\Omega_\mathrm{p}\approx 34.7$ (b) Evolution of the full intracavity intensity $|E(t,\tau)|^2$. (c) Snapshot of the intensity profile around one of the PDCSs at the output of the simulation, $t_\mathrm{f} = 10$. (d) Spectrum of the intracavity field at the simulation output.  (e) Solid curve shows the real part of the intracavity field profile after a numerical filter was used to remove the pump frequencies (shaded region in (d) indicates the filter). Dashed red curve shows the analytical PDCS solutions as a comparison. All the simulation results were obtained from the full Ikeda-like map, except for the thick gray curve in the background of (d) which was obtained using the mean-field Eq.~\eqref{LLN} so as to provide an illustrative comparison between the two models (Ikeda-like map results are shown as the magenta curve on the foreground of (d)).}
 \label{fig2}
\end{figure}

We now test the concept discussed above by performing numerical simulations of the Ikeda map described by Eqs.~\eqref{Seq} and~\eqref{Boundary}. We emphasize that none of the approximations described above (e.g. homogeneity of $E_\pm$) are used -- we directly model the cavity dynamics using the Ikeda map. Our first set of simulations considers a resonator with quintic dispersion profile with $d_2 = -2$ and $d_4 = 0.02$ [see Fig.~\ref{fig2}(a)]. For these simulations, we set $\Delta_+ = \Delta_-= 0$ and $X_\pm \approx 3.58$, such that (in the homogeneous approximation) the stationary intracavity states at the pump frequencies are symmetric with $Y_\pm = 0.685$, thus yielding $|\mu| = 1.37$. It is known that, for such a parametric driving amplitude, PDCSs exist for an effective detuning $\Delta_\mathrm{eff} = 1.2$~\cite{englebert_parametrically_2021}. To obtain this effective detuning for the dispersion profile shown in Fig.~\ref{fig2}(a), we solve Eq.~\eqref{PM} and find that the pump frequency shift $\Omega_\mathrm{p} \approx 34.7$.

Figure~\ref{fig2}(b) shows the evolution of the numerically simulated intracavity intensity profile with an initial condition consisting of two hyperbolic secant pulses with opposite phases. We find that, after a short transient, the intracavity field reaches a steady-state that is indicative of two pulses circulating around the resonator. The pulses sit atop a rapidly oscillating background that is due to the beating between the quasi-homogeneous fields $E_\pm$ at the pump frequencies [Fig.~\ref{fig2}(c)]. Correspondingly, the spectrum of the simulation output [Fig.~\ref{fig2}(d), magenta curve] shows clearly the presence of a hyperbolic secant-shaped feature that sits in between strong quasi-monochromatic components at the pump frequencies. For the sake of completeness, we also show in Fig.~\ref{fig2}(d) as a gray curve the spectrum obtained from simulations that use the mean-field Eq.~\eqref{LLN} -- this result is almost identical with that obtained from the Ikeda-like map, corroborating the congruence of both models. It is also worth noting that, in accordance with PDCS theory [see Eq.~\eqref{PDsol}], there is no significant CW peak at the parametric signal frequency at which the PDCSs are spectrally centred.

To highlight the phase disparity of the pulses, we apply a numerical filter to remove the quasi-monochromatic intracavity components around the pump frequencies, and plot in Fig.~\ref{fig2}(e) the real part of the complex envelope $E(t,\tau)$. Also shown is the corresponding result from the analytical PDCS solutions given by Eq.~\eqref{PDsol}. The simulation results are clearly in excellent agreement with the analytical solution, confirming the possibility of generating PDCSs in a dispersive, pure Kerr resonator.

\begin{figure}[!t]
 \centering
  \includegraphics[width = \columnwidth, clip=true]{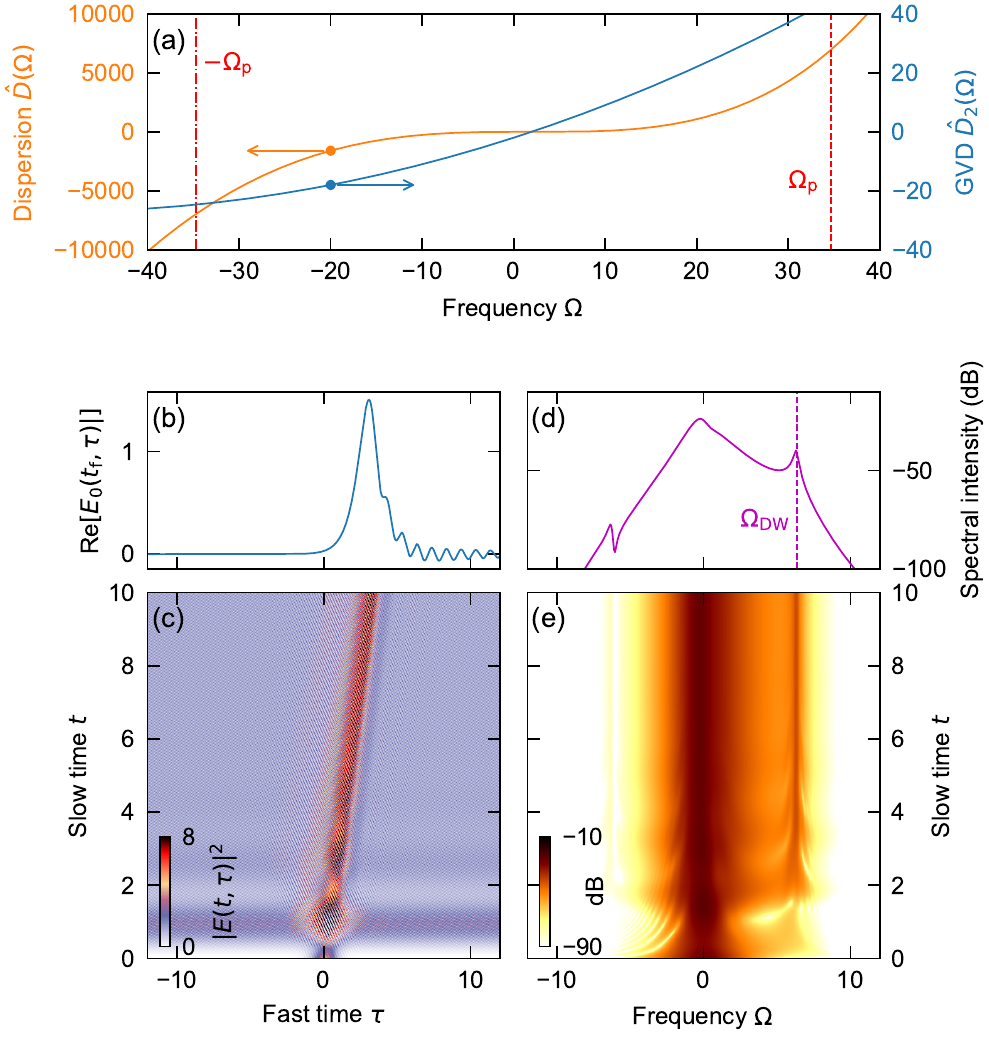}
 \caption{Simulation results illustrating Kerr PDCS generation in the presence of third-order dispersion. (a) Dispersion $\hat{D}(\Omega)$ used in the simulations (orange, left axis) as well as the corresponding group-velocity dispersion $\hat{D}_2(\Omega)$ (blue, right axis). The coefficients $d_2$ and $d_4$ are as in Fig.~\ref{fig2} and $d_3 = 1$. The dash-dotted vertical lines indicate the pump frequency shifts with $\Omega_\mathrm{p}\approx 34.7$. (b) and (d) show temporal and spectral profiles of the PDCS at the simulation output, respectively, while (c) and (e) Shows the corresponding evolutions of the intensity profiles. Note that: (i) fields at the pump frequencies were filtered out for (b) and (ii) the x-axis in (d) and (e) does not extend to the pump frequencies: only frequencies around the PDCS are shown. The dashed vertical line in (d) indicates the theoretically predicted dispersive wave frequency. }
 \label{fig3}
\end{figure}

The results in Fig.~\ref{fig2} were obtained assuming a dispersion profile with no odd-order terms, which may be difficult to realise in practice (at least without an intracavity waveshaper~\cite{runge_pure-quartic_2020, xue_dispersion-less_2021}). To gain more insights, we performed simulations with non-zero third-order dispersion. Figure~\ref{fig3} shows results from simulations with $d_3 = 1$ and with all other parameters (including second and fourth-order dispersion coefficients and the pump frequencies) as in Fig.~\ref{fig2}. As can be seen, even with non-zero third-order dispersion, the Kerr resonator system can support PDCSs. (For the sake of clarity, we considered here an initial condition that results only in a single PDCS, but we have carefully checked that also the PDCS with the opposite phase can be sustained.) As for conventional (externally-driven) Kerr CSs, third-order dispersion causes the solitons to emit dispersive radiation with frequency $\Omega_\mathrm{DW}$ determined by the phase-matching condition $\hat{D}(\Omega_\mathrm{DW})\approx \Delta_0$~\cite{coen_modeling_2013, jang_observation_2014,milian_soliton_2014, brasch_photonic_2016}. This emission in turn causes the solitons to spectrally recoil away from $\Omega_\mathrm{DW}$, which results in the soliton experiencing constant drift in the temporal domain [see Fig.~\ref{fig3}(b)]. It is worth noting that, for the parameters considered in Fig.~\ref{fig3}, the pump frequency $-\Omega_\mathrm{p}$ experiences anomalous group-velocity dispersion; however, the intracavity intensity at that frequency ($Y_- = 0.685$) is below the modulation instability threshold, thus allowing the corresponding field to remain quasi-homogeneous.

\begin{figure}[!b]
 \centering
  \includegraphics[width = \columnwidth, clip=true]{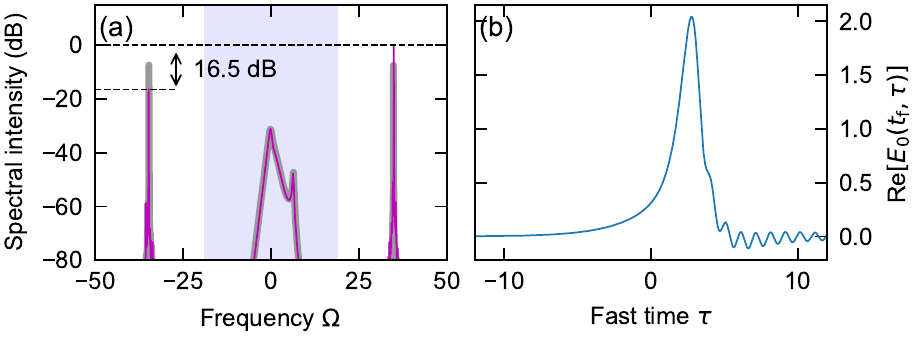}
 \caption{Simulation results with parameters as in Fig.~\ref{fig3} except that the driving intensities are asymmetric with \mbox{$X_+ \approx 117$} and \mbox{$X_- \approx 9.1$}. (a) Magenta curve shows the spectrum at the simulation output. For comparison, the thick gray curve in the background shows the spectrum obtained with identical driving intensities [c.f. Fig.~\ref{fig3}(d)]. (b) PDCS field profile obtained after numerical filtering [shaded region in (a)].}
 \label{fig4}
\end{figure}

\begin{figure}[!t]
 \centering
  \includegraphics[width = \columnwidth, clip=true]{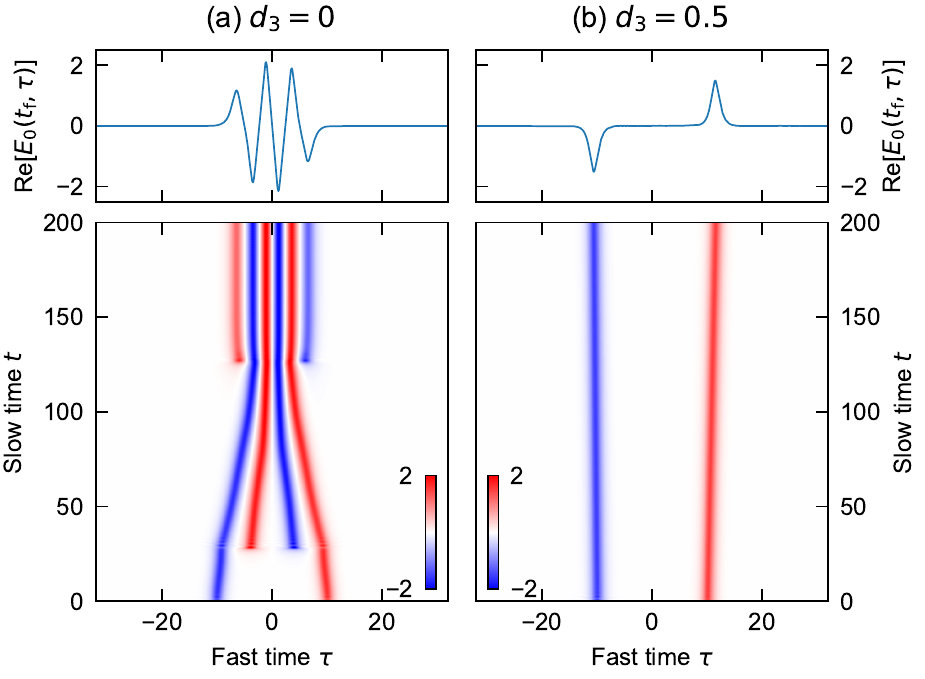}
 \caption{Simulation results illustrating interactions between PDCSs. The pseudo-color plots show the evolution of the real part of the field $E_0(t,\tau)$ that is obtained from the simulated envelope $E(t,\tau)$ by numerically filtering out the intracavity fields at the pump frequencies. The simulation parameters are as in Fig.~\ref{fig2} but with different dispersion profiles: (a) $d_2 = -2$, $d_4 = 0.05$, and (b) $d_2 = -2$, $d_3 = 0.5$, $d_4 = 0.05$. Dispersion coefficients that are not quoted are equal to zero.}
 \label{fig5}
\end{figure}

So far, we have considered parameter configurations that yield identical intracavity intensities $Y_\pm$ at the pump frequencies. This, however, is not a requirement for PDCS existence: it is the product $Y_+Y_-$ that sets the parametric driving amplitude $|\mu| = 2\sqrt{Y_+Y_-}$. Figure~\ref{fig4} shows results that illustrate this point. Here we consider parameters as in Fig.~\ref{fig3} but with the parametric amplitude $|\mu|=1.37$ reached via different driving intensities $X_+ \approx 117$ and $X_- \approx 9.1$, yielding $Y_+ = 4.7$ and $Y_- = 0.1$. The magenta curve in Fig.~\ref{fig4}(a) depicts the spectral intensity at the output of the simulation, showing clearly how the intracavity intensities at the pump frequencies differ by about 16.5~dB. For comparison, also shown as the thick gray curve in the background is the spectrum obtained with identical intracavity intensities $Y_\pm$, extracted from the simulation results shown in Fig.~\ref{fig3}(d). As can be seen, despite the different intracavity intensities $Y_\pm$, the PDCS spectrum is unchanged. Figure~\ref{fig4}(b) shows the field profile around the signal frequency, evidencing a PDCS similar to the one shown in Fig.~\ref{fig3}(b). We note, however, that the PDCS in Fig.~\ref{fig4} differs in phase from the one in Fig.~\ref{fig3} due to the different cross-phase modulations imparted by the fields at the pump frequencies.

As a final set of results, we discuss the impact of the finite temporal walk-off between the interacting fields. Specifically, the intracavity fields at the pump frequencies are strictly homogeneous only when the walk-off between them and the signal field approaches infinity. In the presence of finite walk-off, the interaction of the PDCSs with the fields at the pump frequencies creates localized depletion regions that drift away from the solitons. These depletion regions can create long-range coupling between temporally separated solitons~\cite{jang_ultraweak_2013}, as well as self-interactions in the presence of periodic boundaries. Figure~\ref{fig5}(a) shows results from simulations with parameters as in Fig.~\ref{fig2}, but with a larger fourth-order dispersion coefficient $d_4 = 0.05$ so as to reduce the temporal walk-off between the pump and the signal frequencies (here $\Omega_\mathrm{p} = 22$ with $\hat{D}_1(\pm\Omega_\mathrm{p})=\pm44$). As can be seen, the hyperbolic secant initial conditions first reshape into PDCSs with opposite phases that then experience weak attractive interaction. More complex dynamics emerge when the solitons are sufficiently close to each other: we see how new PDCSs with different phases are spontaneously excited.

The strength of interactions decreases rapidly as the walk-off between the fields increases. This is illustrated in Fig.~\ref{fig5}(b), which shows simulation results with parameters as in Fig.~\ref{fig5}(a) but with the inclusion of a third-order dispersion coefficient $d_3 = 0.5$ that yields $\hat{D}_1(\Omega_\mathrm{p}) = 166$ and $\hat{D}_1(-\Omega_\mathrm{p}) = 76$. (For clarity the results are shown in a reference frame where the PDCSs are stationary.) As can be seen, the inclusion of third-order dispersion effectively suppresses the interactions due to the increased temporal walk-off. Some remaining (repulsive) interactions can still be observed, but we suspect these arise due to the solitons' extended dispersive wave tails rather than pump depletion~\cite{wang_universal_2017}. Finally, while beyond the scope of this work, we expect that the soliton interactions can also be suppressed by considering configurations where the pump fields are non-resonant, such that the intracavity fields at the pump frequencies are fully reset at the beginning of each round trip. In this case, the localized depletion regions that mediate the interactions are removed each round trip, with the fields $E_\pm$ appearing in Eq.~\eqref{PDNLSE} closer to being genuinely homogeneous.  Alternatively, injection of a weak control signal with deterministic phase or amplitude modulations centred around the signal frequency $\omega_0$ could be used to trap the solitons at fixed positions~\cite{englebert_parametrically_2021, jang_temporal_2015}.

\section{Discussion}

Before closing, we discuss opportunities and challenges for experimental realisations. Resonators with appropriate dispersion profiles are already available, as demonstrated by observations of large-frequency shift parametric oscillations in monolithic microresonators~\cite{sayson_widely_2017, sayson_octave-spanning_2019, fujii_third-harmonic_2017, lu_efficient_2019, lu_milliwatt-threshold_2019} and macroscopic optical fiber ring resonators~\cite{li_experimental_2020}. To achieve PDCS generation at $\omega_0$, one must essentially pump these systems at the parametric sidebands at $\omega_\pm = \omega_0\pm \Omega_\mathrm{p}$ under appropriate conditions. One approach would be to pump a fiber ring resonator constructed from fiber with zero-dispersion wavelength close to the 1550~nm telecommunication band~\cite{li_experimental_2020} with two separate external cavity diode lasers with wavelengths on either side of the ZDW. By actively stabilizing the frequencies of both lasers to their respective resonances, we envisage that conditions favourable for PDCS generation can be achieved. We do note that the key role of dispersive walk-off would likely prohibit the use of pulsed pump sources to increase the effective driving strength~\cite{anderson_observations_2016}, but this issue could be mitigated by using an intracavity filter to realise conditions whereby only the parametric signal resonates~\cite{englebert_parametrically_2021}. In this case, however, the pump sources would likely have to be derived from a common CW laser that would then be stabilized to the resonator at the parametric signal frequency (in a manner similar to ref.~\cite{englebert_parametrically_2021}). Alternatively, one could use very short (fiber) resonators with high finesse to enable pure CW pumping (with the short resonator preventing the build-up of stimulated Brillouin scattering~\cite{jang_strong_2012}). An alternative approach could be to integrate a waveshaper inside the resonator so as to engineer custom phase shifts for pump waves with arbitrary frequencies~\cite{runge_pure-quartic_2020, xue_dispersion-less_2021}, and in this way get around the dispersion requirements of the process. Yet a third approach would be to realise PDCSs in a bichromatically pumped microresonator, leveraging the resonator's thermal nonlinearity to passively lock the two pump sources to the cavity. In this case, judicious design of the resonator geometry could enable the dispersion to be engineered to match external cavity diode lasers at hand. 

To conclude, we have theoretically and numerically shown that parametrically-driven cavity solitons can exist in coherently-driven dispersive resonators with pure Kerr nonlinearity. We have discussed the conditions under which such solitons can exist and explored some of their rich dynamics. From a fundamental vantage, our results show that polychromatic-driving can expand the range of localized structures that can be hosted and studied in dispersive Kerr resonators, with a plethora of new dynamics to be explored. Moreover, the bi-phase degeneracy of PDCSs make them attractive candidates for bits in all-optical random number generators~\cite{takesue_10_2016,okawachi_quantum_2016,okawachi_dynamic_2021} or coherent Ising machines~\cite{inagaki_Large-scale_2016, mohseni_ising_2022}, and our work could therefore pave the way for new and possibly improved approaches for realising such applications. Lastly, to the best of our knowledge, our work is the first to present and use an Ikeda-like map model that can account for driving fields associated with multiple frequencies with different detunings. This model can allow for the exploration of effects that cannot be described using the mean-field Lugiato-Lefever formalism~\cite{hansson_frequency_2015,anderson_coexistence_2017}, whilst simultaneously offering benefits in terms of computational efficiency.

\begin{acknowledgments}
We acknowledge support from the Marsden Fund and the Rutherford Discovery Fellowships of the Royal Society of New Zealand. JF acknowledges the CNRS IRP Wall-IN project and the FEDER Solstice project.
\end{acknowledgments}
\appendix
\section{Derivation of the Ikeda map}
\label{apA}
We first present a heuristic derivation of the Ikeda-like map used in our numerical simulations [Eq.~\eqref{Boundary}]. Including the rapid temporal oscillations at $\omega_0$, the intracavity electric field during the $m^\mathrm{th}$ cavity transit is written as $E^{(m)}(z,\tau)\exp[-i\omega_0 T]$, where $\tau$ is time in a co-moving reference frame defined as $\tau = T-z/v_\mathrm{g}$ with $T$ absolute laboratory time, $z$ the coordinate along the waveguide that forms the resonator, $v_\mathrm{g}$ the group velocity of light at $\omega_0$, and $E^{(n)}(z,\tau)$ is the slowly-varying electric field envelope that follows the generalized nonlinear Schr\"odinger equation~\eqref{Seq}. The boundary equation for the full electric field can then be written as
\begin{align}
E^{(m+1)}(0,\tau)e^{-i\omega_0 T} &= \sqrt{1-2\alpha} E^{(m)}(L,\tau)e^{-i\delta_0-i\omega_0 T} \nonumber \\
 &+ \sqrt{\theta_+}E_\mathrm{in,+} e^{-i\omega_+T} \nonumber \\
 &+ \sqrt{\theta_-}E_\mathrm{in,-} e^{-i\omega_-T}, \label{Boundary2}
\end{align}
where $\delta_0 = 2\pi k - \beta(\omega_0) L$ is the linear phase detuning of the reference frequency $\omega_0$ from the closest cavity resonance (with order $k$), and $\omega_\pm$ are the frequencies of the pump fields. Multiplying each side with $\exp[i\omega_0 T]$ and replacing $T = \tau + mL/v_\mathrm{g}= \tau + mt_\mathrm{R}$ where $t_\mathrm{R}$ is the round trip time, yields
\begin{align}
E^{(m+1)}(0,\tau) &= \sqrt{1-2\alpha} E^{(m)}(L,\tau)e^{-i\delta_0} \nonumber \\
 &+ \sqrt{\theta_+}E_\mathrm{in,+} e^{-i\Omega_\mathrm{p}\tau + i(\omega_0-\omega_+)mt_\mathrm{R}} \nonumber \\
 &+ \sqrt{\theta_-}E_\mathrm{in,-} e^{i\Omega_\mathrm{p}\tau + i(\omega_0-\omega_-)mt_\mathrm{R}}, \label{Boundary3}
\end{align}
Note that, in the above formulation, the co-moving time variable $\tau$ should be understood as the ``fast time'' that describes the envelope of the intracavity electric field over a single round trip, i.e., the distribution of the envelope within the resonator~\cite{pasquazi_micro-combs_2018}. As such, the $\tau$ variable spans a single round trip time of the resonator, and the intracavity envelope must obey periodic boundaries within that range. These conditions stipulate that the frequency variable $\Omega_\mathrm{p} = 2\pi p \times \text{FSR}$, where $p$ is a positive integer and $\text{FSR}$ is the free-spectral range of the resonator. The fact that the frequency difference $(\omega_0-\omega_\pm)/(2\pi)$ may not, in general, be an integer multiple of the FSR is captured by the additional phase shifts accumulated by the driving fields with respect to the intracavity field from round trip to round trip.

To link the frequency differences $\omega_0-\omega_\pm$ to the respective phase detunings, we first recall that the phase detuning of a driving field with frequency $\omega$ from a cavity resonance at $\omega'$ obeys $\delta\approx (\omega'-\omega)t_\mathrm{R}$. We can thus write
\begin{equation}
\omega_0-\omega_\pm \approx  \omega'_0-\frac{\delta_0}{t_\mathrm{R}} -  \omega'_{\pm} + \frac{\delta_\pm}{t_\mathrm{R}}, \label{freqdif}
\end{equation}
where the frequency variables with (without) apostrophes refer to resonance (pump) frequencies. We next write the resonance frequencies as $\omega'_q = \omega'_0 + q \zeta_1 + \hat{D}_\mathrm{int}(q)$, where $\zeta_1 = 2\pi\text{FSR}$ with $\text{FSR} = t^{-1}_\mathrm{R}$ the free-spectral range of the cavity (at $\omega'_0$), $q$ an integer that represents the mode index (with $\omega_0$ corresponding to $q = 0$), and the integrated dispersion
\begin{equation}
\hat{D}_\mathrm{int}(q) = \sum_{k\geq 2} \frac{\zeta_k}{k!}q^k,
\label{Dint}
\end{equation}
where $\zeta_k$ are the expansion coefficients. Assuming that the resonance frequencies $\omega'_\pm$ are associated with indices $\pm p$ (with $p>0$), respectively, we can write Eq.~\eqref{freqdif} as
\begin{equation}
\omega_0-\omega_\pm \approx  -\frac{\delta_0}{t_\mathrm{R}} \mp p \zeta_1 - \hat{D}_\mathrm{int}(\pm p) + \frac{\delta_\pm}{t_\mathrm{R}}. \label{freqdif2}
\end{equation}
The second term on the right-hand-side of Eq.~\eqref{freqdif2} can be ignored, as it yields an integer multiple of $2\pi$ when used in Eq.~\eqref{Boundary3}. Next, we use the fact~\cite{pasquazi_micro-combs_2018} that the coefficients $\zeta_k$ with $k\geq 2$ can be linked to the Taylor series expansion coefficients of the propagation constant $\beta(\omega)$ viz. $\zeta_k \approx - \zeta_1^kL\beta_k/t_\mathrm{R}$. This allows us to write the integrated dispersion corresponding to the resonance frequencies $\omega'_\pm$ as
\begin{align}
\hat{D}_\mathrm{int}(\pm p) &\approx \sum_{k\geq 2} -\frac{\zeta_1^kL\beta_k}{t_\mathrm{R}k!}(\pm p)^k, \\
& = -\frac{L}{t_\mathrm{R}}\sum_{k\geq 2} \frac{\beta_k}{k!}(\pm \zeta_1 p)^k.
\label{Dint2}
\end{align}
Using Eq.~\eqref{Dint2} in Eq.~\eqref{freqdif2} and substituting the latter into Eq.~\eqref{Boundary3} yields the Ikeda-like map described by Eq.~\eqref{Boundary} with coefficients $b_\pm$ as defined in Eq.~\eqref{bs}, and the pump frequency shift $\Omega_\mathrm{p} = \zeta_1 p = 2\pi p \times \text{FSR}$. We also note that the map can be straightforwardly extended to include arbitrarily many driving fields.

\section{Signal detuning and desynchronization}
\label{apA2}
To derive the relationship between the parametric signal detuning $\delta_0$ and the pump detunings $\delta_\pm$ (i.e., Eq.~\eqref{d0}), we write out the parametric signal frequency as
\begin{equation}
\omega_0 = \frac{\omega_+ + \omega_-}{2} = \frac{\omega_+' - \Delta\omega_+ + \omega_-'- \Delta\omega_-}{2},
\end{equation}
where $\omega_\pm'$ are the resonance frequencies closest to the pump frequencies and $\Delta\omega_\pm$ are the angular frequency detuning of the pump frequencies from those resonances. Substituting $\Delta\omega_\pm = \delta_\pm/t_\mathrm{R}$ and expanding the pump resonance frequencies as $\omega_\pm' = \omega_0' \pm p\zeta_1 + \hat{D}_\mathrm{int}(\pm p)$ yields
\begin{equation}
\omega_0 = \omega_0' - \frac{\delta_+ + \delta_-}{2t_\mathrm{R}} + \frac{\hat{D}_\mathrm{int}(p) + \hat{D}_\mathrm{int}(-p)}{2}.
\end{equation}
Then using Eq.~\eqref{Dint2} and rearranging, we obtain
\begin{align}
\delta_0 &= (\omega_0'-\omega_0)t_\mathrm{R} \nonumber\\
&= \frac{\delta_+ + \delta_- +  L[\hat{D}_\mathrm{S}(\Omega_\mathrm{p})+\hat{D}_\mathrm{S}(-\Omega_\mathrm{p})]}{2}.
\end{align}
This is Eq.~\eqref{d0}.

Proceeding in a similar fashion, we next examine under what conditions the sinusoidal modulation associated with the pump fields is synchronous with the resonator free-spectral range. To this end, we consider the difference between the pump frequencies:
\begin{align}
\omega_+-\omega_- &= \omega_+' - \Delta\omega_+-\omega_-' +\Delta\omega_- \\
& =  2p\zeta_1 + \hat{D}_\mathrm{int}(p) - \hat{D}_\mathrm{int}(-p) + \frac{\delta_-}{t_\mathrm{R}} - \frac{\delta_+}{t_\mathrm{R}}.
\end{align}
Rearranging and using Eq.~\eqref{Dint2} yields
\begin{equation}
\omega_+-\omega_- = 2p\zeta_1 - \frac{2}{t_\mathrm{R}}\left(\frac{\delta_+ - \delta_- + L[\hat{D}_\mathrm{S}(\Omega_\mathrm{p}) - \hat{D}_\mathrm{S}(-\Omega_\mathrm{p})]}{2}  \right).
\end{equation}
We can recognise the term inside parentheses as the coefficient $b$ introduced in Eq.~\eqref{bcoef}. Thus, we see that when $b = 0$, the modulation frequency $(\omega_+-\omega_-)/(2\pi)$ is equal to an integer multiple of the cavity FSR, i.e., the modulation is synchronous with the cavity round trip time. In contrast, when $b\neq 0$, the modulation is de-synchronized. The coefficient $a$ introduced in Eq.~\eqref{acoef} is the normalized version of $b$, and it is therefore not surprising that this term appears as a de-synchronization in Eq.~\eqref{cosdrive}.

\section{Normalized Lugiato-Lefever equation}
\label{apB}
Under the assumption that the intracavity envelope $E^{(m)}(z,\tau)$ evolves slowly over a single round trip (i.e., the cavity has a high finesse, and the linear and nonlinear phase shifts are all small), the Ikeda-like map described by Eqs.~\eqref{Seq} and~\eqref{Boundary}  can be averaged into a single mean-field equation. The derivation is well-known~\cite{haelterman_dissipative_1992}, proceeding by integrating Eq.~\eqref{Seq} using a single step of the forward Euler method to obtain $E^{(m)}(L,\tau)$, which is then substituted into Eq.~\eqref{Boundary}. After linearizing with respect to $\delta_0$ and $\alpha$ and introducing the \emph{slow} time variable $t = mt_\mathrm{R}$ (such that the round trip index $m = t/t_\mathrm{R}$), one obtains:
\begin{align}
t_\mathrm{R}\frac{\partial E(t,\tau)}{\partial t} &= \left[-\alpha + i(\gamma L |E|^2 -\delta_0)+ iL\hat{D}_\mathrm{S}\left(i\frac{\partial}{\partial\tau}\right)\right]E \nonumber \\
 &+ \sqrt{\theta_+}E_\mathrm{in,+} e^{-i\Omega_\mathrm{p}\tau + ib_+t/t_\mathrm{R}} \nonumber \\
 &+ \sqrt{\theta_-}E_\mathrm{in,-} e^{i\Omega_\mathrm{p}\tau + ib_-t/t_\mathrm{R}}. \label{LLE}
\end{align}
To obtain the normalized Eq.~\eqref{LLN}, we first introduce the variable transformations $\tau\rightarrow \tau \sqrt{2\alpha/(|\beta_2|L)}$, $t\rightarrow \alpha t/t_\mathrm{R}$, $\Omega_\mathrm{p}\rightarrow \Omega_\mathrm{p}\sqrt{|\beta_2|L/(2\alpha)}$ and $E\rightarrow E \sqrt{\gamma L/\alpha}$, yielding
\begin{align}
\frac{\partial E(t,\tau)}{\partial t} &= \left[-1 + i(|E|^2 -\Delta_0)+ i\hat{D}\left(i\frac{\partial}{\partial\tau}\right)\right]E \label{LLN2} \\
 &+ S_+ e^{-i\Omega_\mathrm{p}\tau + ia_+t} \nonumber + S_- e^{i\Omega_\mathrm{p}\tau + ia_-t},  \nonumber
\end{align}
where $S_\pm = E_\mathrm{in,\pm} \sqrt{\gamma L \theta_\pm/\alpha^3}$, and $\Delta_0 = \delta_0/\alpha$. The normalized dispersion operator $\hat{D}$ is defined as
\begin{equation}
\hat{D}\left(i\frac{\partial}{\partial\tau}\right) = \sum_{k\geq 2} \frac{d_k}{k!}\left(i\frac{\partial}{\partial\tau}\right)^k,
\label{dispop2}
\end{equation}
where the normalized dispersion coefficients are given by
\begin{equation}
d_k = \frac{\beta_k L}{\alpha}\left(\frac{2\alpha}{|\beta_2| L}\right)^{k/2}.
\end{equation}
Finally, the coefficients $a_\pm = \Delta_\pm - \Delta_0 + \hat{D}(\pm \Omega_\mathrm{p})$, where $\Delta_\pm = \delta_\pm/\alpha$ are the normalized detunings of the external driving fields. Note that, for the particular configuration considered in our work, where the signal frequency $\omega_0$ is strictly linked to the pump frequencies $\omega_\pm$ via $\omega_0 = (\omega_+ + \omega_-)/2$, the coefficients $a_\pm = \pm a$, where $a$ is defined by Eq.~\eqref{acoef}.

\section{Homogeneous states at pump frequencies}
\label{apC}
When the ansatz described by Eq.~\eqref{anz} is substituted into Eq.~\eqref{LLN} and terms oscillating at different frequencies isolated, one obtains the following evolution equations for the homogeneous intracavity fields $E_\pm(t)$:
\begin{align}
\frac{\mathrm{d} E_\pm(t)}{\mathrm{d} t} &=\left[-1 + i(|E_\pm|^2 + 2|E_\mp|^2 - \Delta_\pm)\right]E_\pm + S_\pm, \label{Epm}
\end{align}
where $\Delta_\pm = \Delta_0 + a_\pm - \hat{D}(\pm\Omega_\mathrm{p})$. Note that, in deriving Eqs.~\eqref{Epm} we have ignored the coupling between the intracavity fields at the pump frequencies and the PDCS at the signal frequency, which is justified when the walk-off between the fields is large.  The steady-state ($\mathrm{d} E_\pm/\mathrm{d}t = 0$) solutions of Eqs.~\eqref{Epm} satisfy
\begin{equation}
E_\pm = S_\pm/[1-i(Y_\pm + 2Y_\mp - \Delta_\pm)],
\end{equation}
where the intensities $Y_\pm = |E_\pm|^2$ satisfy Eqs.~\eqref{Xs}.

\end{document}